\documentclass[conference]{IEEEtran}
\usepackage{graphicx}
\usepackage{caption}
\usepackage{subcaption}

\usepackage{longtable}
\usepackage{stfloats}
\usepackage{graphicx}

\usepackage[draft]{hyperref}

\usepackage{ifpdf}
\usepackage[T1]{fontenc}
\usepackage{amssymb}
\usepackage{cite}

\ifCLASSINFOpdf
   \usepackage{graphicx}
	\usepackage{epstopdf}
   \graphicspath{{images/}}   
   \graphicspath{{../pdf/}{../jpeg/}{../eps/}}
   \DeclareGraphicsExtensions{.pdf,.jpeg,.png}
\else
   \usepackage[dvipdfm]{graphicx}
   \usepackage{bmpsize}
   \graphicspath{{../eps/}}
   \DeclareGraphicsExtensions{.eps}
\fi
\graphicspath{{images/}}

\usepackage[cmex10]{amsmath}
\usepackage{multirow}

\usepackage{authblk}

\begin{document}

\title{Design of a Highly Reliable Wireless Module for Ultra-Low-Latency Short Range Applications}

\author[1]{Raja Sattiraju}
\author[2]{Jasper Siemons}
\author[3]{Mohammad Soliman}
\author[4]{Wasim Alshrafi}
\author[3]{Fabian Rein}
\author[1]{Hans D. Schotten}

\affil[1]{ University of Kaiserslautern \    \url{(sattiraju, schotten)@eit.uni-kl.de}}
\affil[2]{IMST GmbH \       \url{siemons@imst.de}}
\affil[3]{German Aerospace Center (DLR) \ \url{(mohammad.soliman, fabian.rein)@dlr.de}}
\affil[4]{RWTH Aachen University \    \url{alshrafi@ihf.rwth-aachen.de}}

\maketitle

\begin{abstract}
Current radio systems are currently optimized for capacity and range. However, certain applications of wireless systems require fast and reliable communication over short distances. The challenge of these systems is to communicate with a minimum time delay (latency) while at the same time being very reliable and resilient to interference. This paper describes the concept and the proposed abstract architecture of a wireless technology that allows highly reliable and ultra low latency transmission of data between moving units over a few meters, the applications of which can be found in multitude of domains such as Train Communicaion Networks (TCNs), Truck/Tractor - Trailer Communication, Platooning, and in smart industry in the form of co-ordinating machines. The paper also describes the set of novelties that were planned to be realized as part of the final demo hardware.
\end{abstract}

\section{Introduction}
{\let\thefootnote\relax\footnote{This is a preprint, the full paper is published in Proceedings of 13th International Workshop on Factory Communication Systems (IEEE WFCS 2017), \copyright 2017 IEEE. Personal use of this material is permitted. However, permission to use this material for any other purposes must be obtained from the IEEE by sending a request to pubs-permissions@ieee.org.}}

Ultra-Low Latency and highly reliable short range communications are being increasingly investigated by many researchers due to their potential applications in various fields such as In-Train Communication Networks (TCN) Vehicle-to-Vehicle (V2V) communications, Wireless Sensor Networks (WSNs), Factory Automation etc. These applications typically demand a level of \textit{reliability} and \textit{timeliness} that current wireless communication systems typically are not able to guarantee\cite{Schotten2014}. The project SBDist was initiated under the framework of Industrie 4.0 by the German Federal Ministry of Education and Research (BMBF) to investigate and design a Low Latency Highly Reliable short range wireless device that is resilient to interference. The application domains for such a device includes Train Communication Networks (TCNs), automotive platooning and smart factory concepts such as Augmented Reality (AR) based maintenance, Co-ordinating machines etc. \

Owing to reasons such as robustness and safety, the trains currently are equipped with cable based networking modems (Ex, UIC Bus) that offer moderate data rates. However with challenges such as increased traffic, ensuring passengers safety and security during their journey, improving travel comfort, providing real time multimedia information and access to social networks in stations or in motion, train operators are searching for flexible, scalable and low cost networking solutions. \

A wireless solution for Wired Train Bus (WTB) offers advantages such as lowering hardware costs and support for new applications. From the implementation perspective, the SBDist module hence must support a diverse set of requirements ranging from applications such as Train Control that require ultra-low latency communication (in the order of $\mu$sec) to multimedia rich applications that are data intensive but have tolerable latency constraints including applications requiring Real-Time Ranging \& Localization (RTLS). \

\section{Use Cases, Requirement Profiles and Service Profiles}

The Use Cases were derived for the train domain,  commercial vehicle domain and the smart industry domain \cite{SBDist2016a}. These are denoted by Use Case-Train (UCT), Use Case-Vehicle (UCV) and Use Case-Industy (UCI) accordingly and are listed as follows. 

\subsection{Train Domain}
For the train domain the following UCs are proposed. These UCs have Hard Real Time requirements (HRTs).
\subsubsection{Train Control and Management Systems (HRT-UCT1)}
TCMS operations are carried out via the Train Communication Network (TCN) which in turn consists of two hierarchical levels - the Wire Train Bus and the Multifunction Vehicle Bus (MVB). The WTB interconnects two or more rail vehicles whereas the MVB is used for intra-vehicular communications \cite{Kirrmann2001a}.

\subsubsection{Automatic Train Pairing (HRT-UCT2)}
When coupling two trains in a station, the rear train usually is stopped in safe distance of several meters behind the waiting front train. The driver then has to approach the front train with very low velocity and stops manually as soon as the couplings lock mechanically. Both units can now initiate an electronic handshaking process to finalize the software-wise pairing. The entire train coupling and pairing process including stop and slow approach takes about one minute. During this time the passenger doors cannot be opened yet, so the overall stop time spent in the station is increased. The SBDist module can be used to establish communication between both trains before mechanical coupling. By measuring precise distance information, the rear train can automatically adjust and control the approaching speed and stop at the exact location for the couplings to lock mechanically.\

%

\subsection{Commercial Vehicle Domain}
For the commercial vehicle domain, the following UCs were proposed.

\subsubsection{Automotive Platooning (HRT-UCV1)}
In order to safely operate a platooning system, the controller system needs frequent and timely information about vehicles in a platoon (assumed to be 10 Hz in general) to enable shorter distances among vehicles.  In \cite{Ploeg2011a}, the authors presented a system wherein the distance that can be maintained is speed dependent. If wireless vehicular communication is used to disseminate the speed information to the direct follower, the headway time can be reduced to 0.5-0.7 seconds compared with the radar only system requiring 1 - 1.5 seconds to avoid compromising the stability of the system.\

\subsubsection{Trailer Control (HRT-UCV2)}
The wired communication systems that exist between a tractor and a trailer today use traditional wire coupling for information transfer (Intra-vehicular communication). This communication configuration is less than desirable for a number of reasons since the wire coupling was originally designed for simple functionality and cannot handle complex data transfer. Specifically, the wire coupling was designed for functions such as transmitting power to the tail lights and providing signals to the brake lights and blinkers on the trailer. Because the wire coupling was not intended for complex data transfer, the wires are not adequately protected and occurrences of interference are typical, thus affecting the communication in the tractor-trailer environment. Efficiency and performance of tractor-trailer distribution schemes are hindered by the lack of available communication. For instance, where a tractor mistakenly attaches the wrong trailer, there is no effective way to communicate the mistake to the driver and, thus, the wrong cargo is transported, wasting time and fuel and perhaps compromising the quality of the cargo.\

UC's dealing with Augmented Reality, Teleoperation and Telepresence were also defined for the Industrial Domain in addition to other lower priority UCs with Soft and Weak Real-Time requirements (SRTs and WRTs). A detailed list of UC's can be found in the Requirement Specification document\cite{SBDist2016a}.


\subsection{Requirements Profiles (RPs)}
The proposed UCs have varying performance requirements with respect to factors such as maximum tolerable latency, required PER (Packet Error Rate) \& data rate etc. Hence, they were classified into four categories of Requirement Profiles (RPs) with each RP having its own set of performance requirements as shown in Table.~\ref{RP-Table}

\begin{table}[]
\centering
\caption{Requirement Profiles}
\label{RP-Table}
\begin{tabular}{lllll}
\textbf{Profile}       & \textbf{RP1}        & \textbf{RP2}        & \textbf{RP3}         & \textbf{RP4}        \\
\textbf{Latency}       & \textless 50 us     & \textless 1 ms      & \textless 10 ms      & \textless 100 ms    \\
\textbf{Bitrate}       & \textless 2 Mbit/s  & \textless 5 Mbit/s  & \textless 100 Mbit/s & \textless 1 Gbit/s  \\
\textbf{LOS}           & Yes                 & Yes                 & Yes                  & Yes                 \\
\textbf{PER}           & \textless $10^{-9}$ & \textless $10^{-9}$ & \textless $10^{-4}$  & \textless $10^{-4}$ \\
\textbf{Distance (cm)} & 20 - 50             & 20 - 200            & 20 - 500             & 20 - 200            \\
\textbf{P2P Link}      & Yes                 & Yes                 & Yes                  & Yes                 \\
\textbf{Security}      & High                & High                & Medium               & Medium              \\
\textbf{HW Redundancy} & Yes                 & Yes                 & No                   & No                 
\end{tabular}
\end{table}

\subsection{Service Profiles (SPs)}
A service profile can be defined as a customized configuration for a logical transmission channel or in other words, the configurations which the modem can achieve. These are derived from the Requirement Profiles (RPs) with each SP supporting a certain Quality of Service (QoS) as specified by the RP. Each SBDist modem can transfer multiple logical channels in parallel with each logical channel having a unique SP differentiated by properties such as Maximum Bitrate ($R_b$),Spreading Factor (SF) etc. Some example SPs are shown in Table.~\ref{SP-Table}. The required modem resources ($R_{req}$) in terms of modem capacity of $C$ Mbits/s can then be calculated as $R_{req} = R_b/C*SF$.\


\begin{table}[]
\centering
\caption{Service Profiles}
\label{SP-Table}
\begin{tabular}{llll}
\multicolumn{1}{c}{\textbf{Service Profile}} & \multicolumn{1}{c}{\textbf{Type}} & \multicolumn{1}{c}{\textbf{$R_b$(Mb/s)}} & \multicolumn{1}{c}{\textbf{SF}} \\
\textbf{SP1}                                 & Improved Robustness               & 25                                       & 8                               \\
\textbf{SP2}                                 & Normal Robustness                 & 200                                      & 2                               \\
\textbf{SP3}                                 & High Data Rate                    & 1000                                     & 1                              
\end{tabular}
\end{table}

\section{Overall SBDist Architecture}
\begin{figure}[h]
\centering
\includegraphics[width=0.48\textwidth]{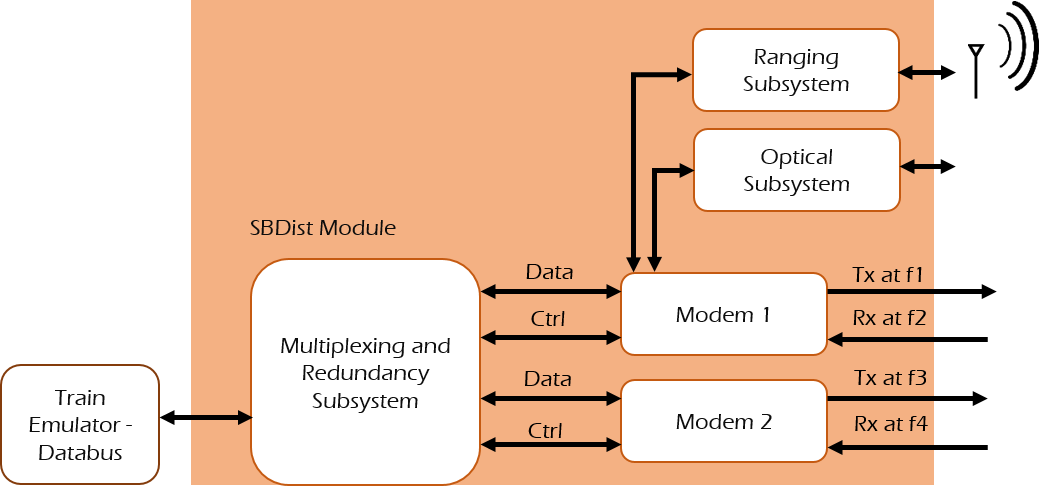}
\caption{SBDist Module Overall Architecture}
\label{fig_architecture}
\end{figure}

The SBDist system consists of two SBDist modules (Two transceivers) that are individually decomposed into the following subsystems as shown in Fig.~\ref{fig_architecture}

\begin{itemize}
\item Ranging subsystem to determine the distance between the two trains and also for low data rate communication
\item Optical Subsystem for increasing the communication performance, reliability and also to serve as a redundant system if the RF system fails due to jamming
\item Multiplexing subsystem for interfacing between the application domain (UIC bus) and the modem
\item Modem Subsystem to modulate and code the received bits from the Mux-subsytem for analog conversion and mixing it with the carrier frequency
\item Antenna Subsystem
\end{itemize}

Further implementation details of the sub-systems can be found in \cite{SBDist2016b}\cite{SBDist2016c}.\

\section{Novelty}

In order to guarantee the performance requirements of the WTB (classified under RP-1), tailormade design solutions are necessary. The following points outline the novelty of the SBDist module

\subsection{Transmission Redundancy with a novel multiplexer and for ultra-high reliability}
The mux subsystem can be separated into the application interface and a control unit. The application interface provides the connection to the Wire Train Bus (WTB), the UIC signals and a user defined media, which can be a 100Base-TX Ethernet interface . It is responsible for signal conditioning and for converting signals from analogue to digital domain and vice versa.\

The data (TCMS and other) is fed into the multiplexer of the SBDist module via the UIC and WTB (or an emulator). The control unit multiplexes/demultiplexes the digital data according to the service profile (based on the RP) of the associated interface into/from the data stream. It uses 2 modems to transfer the data in a redundant/distributive manner depending upon the reliability requirement.\


\subsection{Dedicated logical data streams depending upon maximum allowed deadline to support ultra-low latencies}

\begin{figure}[h!]
\centering
\includegraphics[width=0.48\textwidth]{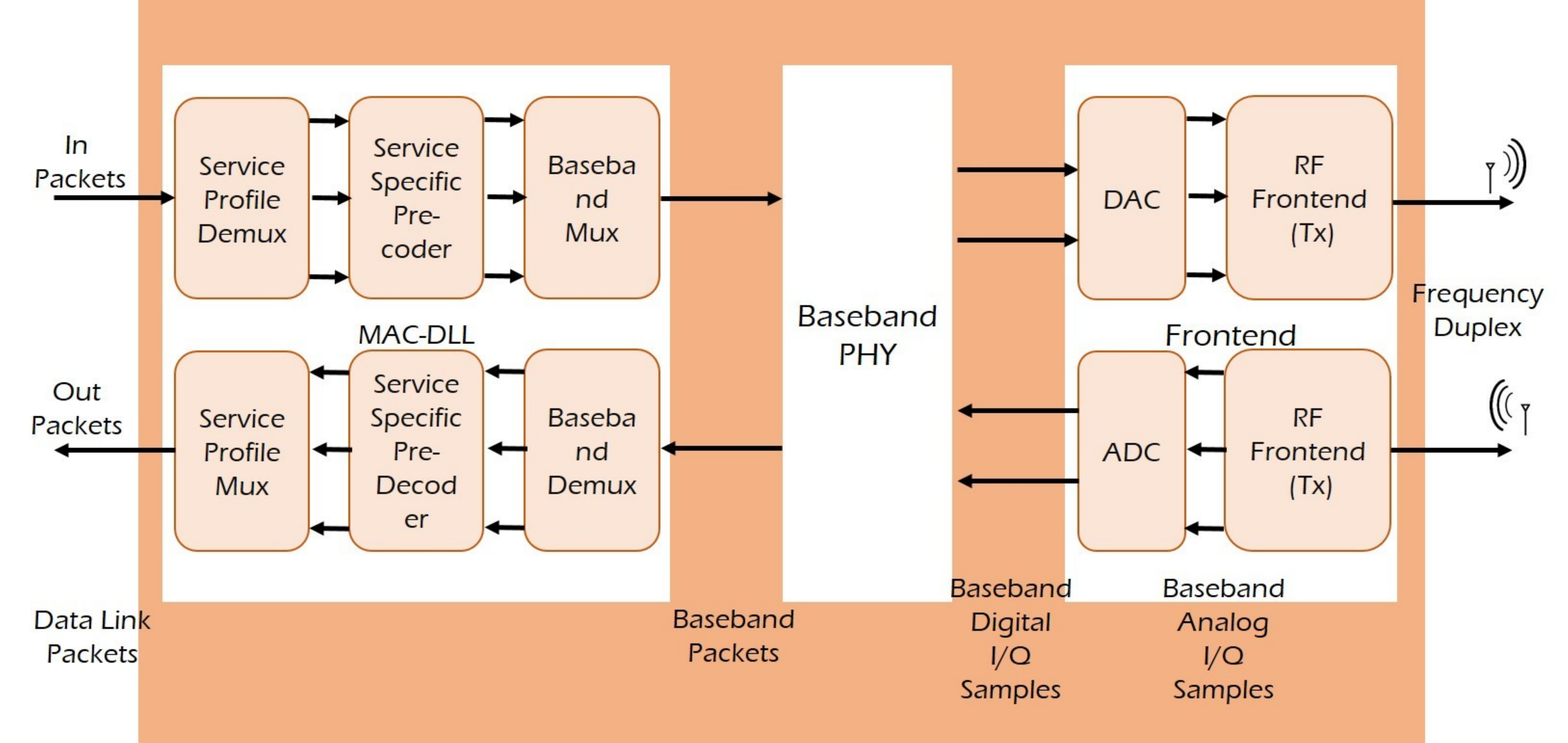}
\caption{SBDist Modem Subsystem}
\label{fig_modem}
\end{figure}

After signal conditioning at the multiplexer, the data is transmitted wirelessly by the modem subsystem which consists of three blocks: a MAC-DLL block, a Baseband block and a Frontend block as shown in Fig.~\ref{fig_modem}.

The MAC-DLL block interfaces externally with the mux-subsystem and internally with the Baseband block.The transmitting part of this block must interpret the data link packets (i.e determine for example the service profile) received from the mux subsystem. Each logical channel must be applied to the corresponding part of the pre-coder block. The pre-coded data is applied to the baseband mux which combines the data into a baseband packet for the baseband block. The receiving part of the MAC-DLL block must do the opposite.\

The design of the baseband focuses on low latency data transmission in the order of 50 $\mu$s with small to moderate data rates up to a few hundred Mbps and very low packet error rates below $10^{-9}$. Only peer-to-peer links are supported allowing for a Frequency Division Duplex (FDD) implementation with continuous data transmissions in transmit and receive direction in order to minimise latency.\

The low packet error rate requires robust and efficient frequency, phase and time synchronization in combination with powerful error control coding. Due to the continuous data transmissions in the FDD mode, synchronization only needs to be acquired once. Hence, acquisition time and performance are of minor relevance, while the focus will be on robust and efficient symbol and phase tracking algorithms. The remaining bit errors will be removed by the error control code. It is well known from theory, that the word error rate (WER) depends on the code word length. Nevertheless a code word length above 1000 bits already exploits most of the available coding gain. In combination with a coded data rate in the order of 500 Mbps the resulting code word duration is a few $\mu s$, which is well below the target latency value.\

\subsection{Simultaneous Ranging and Communication / Dedicated Ranging subsystem for RTLS}
Ranging refers to estimation of distance (range) between two communicating nodes \cite{Kraemer2009}. Simultaneous ranging and communications implies using the same waveform for decoding the data (at the receiver) and also process the back scattered signal (at the transmitter) using traditional radar techniques.\


\subsubsection{Time of Flight (ToF) Ranging}
Using a highly accurate clock, distance between two nodes can be calculated by means of exchanging data packets embedded with the ToF information. This can be implemented by integrating the ToF packet to the legacy data transmission or by integrating existing Commercial Off The Shelf (COTS) components into the ranging subsystem. Examples include Time Domain P440, Decawave EVB 1000 etc.\

\subsubsection{Doppler/Range Radar Processing}
An alternate way to measure the distance between two nodes is to use the Doppler/Range processing techniques from the radar domain by operating the SBDist modem in a full duplex mode and listening to its own transmissions. This concept of combined usage as a radar and communications device has been proposed by \cite{Wiesbeck2001} with theoretical aspects explained in \cite{Braun2014}. The distance can be measured by comparing the phase/frequency shift of the back scattered signal to that of the original signal. However,the receiving chain would have to process the (weak) reflected signal on the same frequency and concurrently to the original (strong) transmission. As in full duplex communication, this is difficult due to self-interference and requires novel signal processing techniques to overcome this problem.\

The usage of Ultra-Wide Band(UWB) frequencies (with bandwidth $\geq$ 500 MHz) for the SBDist system provides an unique opportunity to detect and range obstacles at a much higher granularity due to the short symbol times. This property is exploited to construct a supervised learning hypothesis and using the existing machine learning algorithms to detect and range the obstacles\cite{Sattiraju2017a}. The advantage of this approach is that it can be directly applied on the raw waveform without the need for any signal processing.

\subsection{Single Carrier based Air Interface Technologies}


Two baseband concept designs are proposed. The first concept is a Single Carrier (SC) concept which focuses on the UCT1 scenario in the train domain. The scenario is characterized by a quasi-static relative position between transmit and receive antennas. The novel circular polarized directive antenna array design combined with the quasi-static channel result into a significant reduction of multipath impact. The resulting AWGN like channel enables the use of a single carrier system concept. Due to the moderate bit rate requirements, only low order modulation schemes like BPSK and QPSK will be needed. As a result, the peak-to-average power ratio (PAPR) in the SC concept will be much smaller than in a comparable multi-carrier system, significantly reducing the linearity requirements in the RF part. The design of the transmitter and receiver allows the system to reach the ultra-low latency requirements.\

\begin{figure}[h!]
\centering
\includegraphics[width=0.48\textwidth]{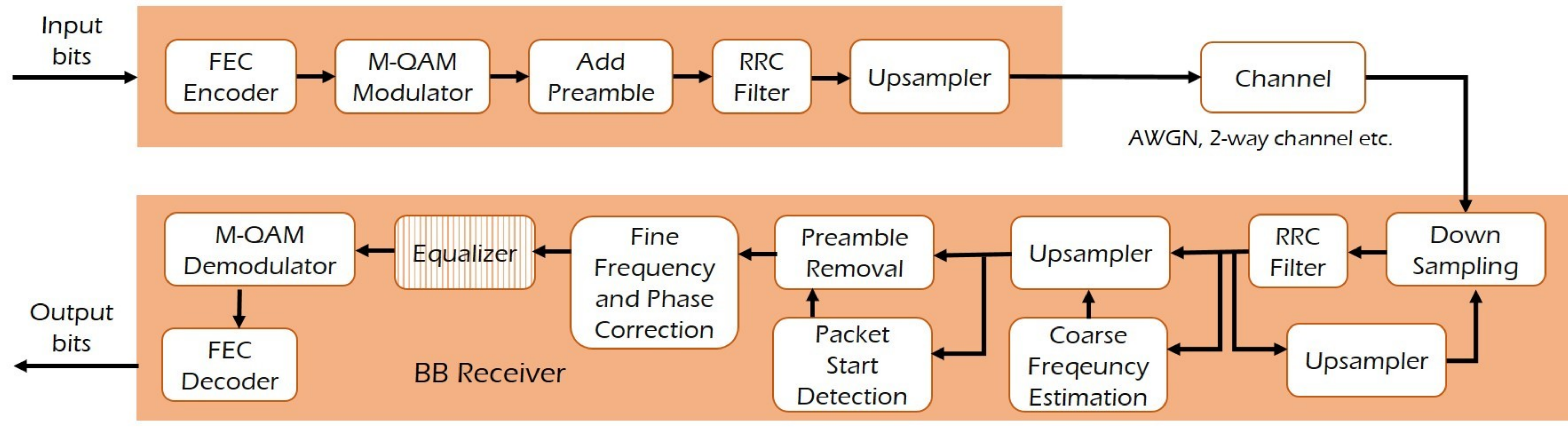}
\caption{SC-FDD with TD equalizer}
\label{fig_fdma_td}
\end{figure}

\begin{figure}[h!]
\centering
\includegraphics[width=0.48\textwidth]{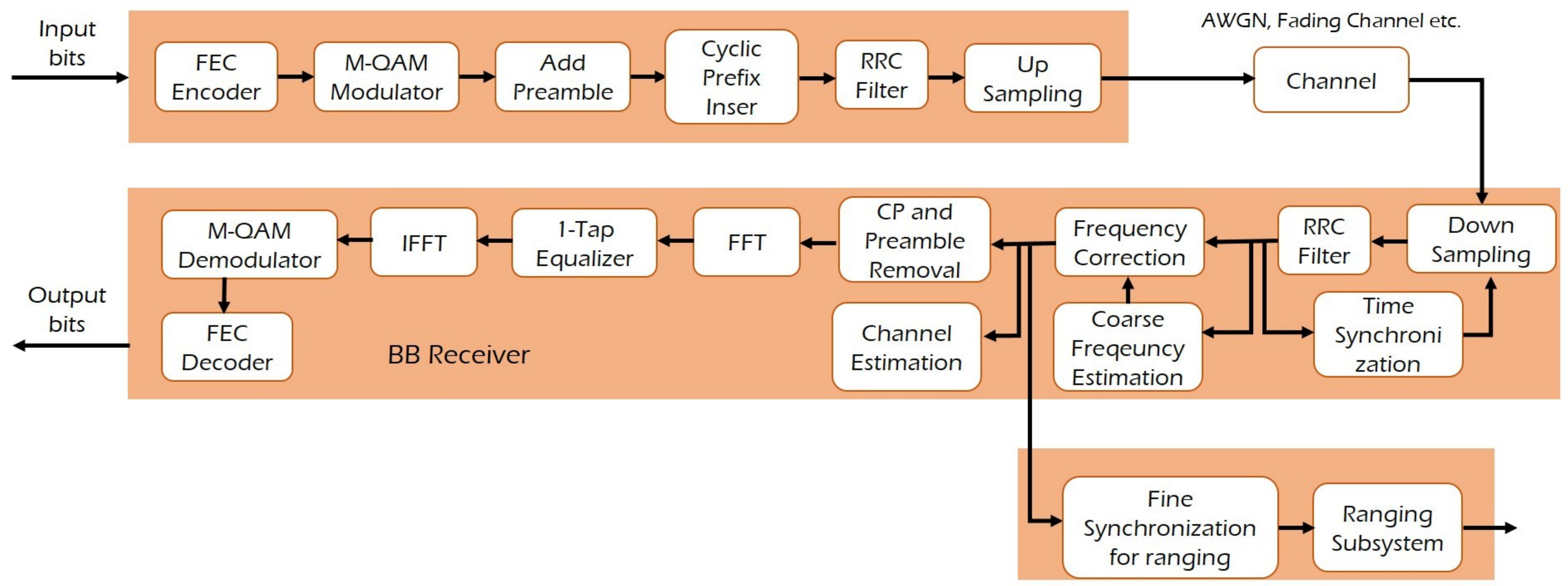}
\caption{SC-FDD with FD equalizer}
\label{fig_fdma_fd}
\end{figure} 

The second concept focuses in more challenging channel requirements covering UCT1-4 scenarios in the train domain. The concept is based on the assumption that the channel might be affected by multipath components due to the wider beam-width of the transmitting and receiving antennas. The metallic body of the train and the coupling area, vibration caused by the train and the surrounding environment are also assumed to contribute in the variation of the channel. A Single Carrier with Frequency Domain Equalization (SC-FDE) system provides simple transmitter architecture with the advantage of being resilient to multipath propagation while maintaining low PAPR property compared to Orthogonal Frequency Division Multiplexing (OFDM). In addition, Cyclic Prefix (CP) is added to the transmit packets to mitigate inter-symbol interference (ISI). The transmitter has a SC system structure; allowing us to use the SC-FDE concept as a backup or an alternative system together with the other SC system. The system also uses the concepts of fine time and frequency synchronization to enable ranging applications. The block diagrams for the two concept baseband system designs are shown in Fig.~\ref{fig_fdma_td} and Fig.~\ref{fig_fdma_fd}.\

\subsection{Circular Polarized Antenna with high directivity}
The baseband concept focuses on the UCT 1-4 scenarios in the train domain. For the train domain use cases, the channel impulse response has multipath components in addition to the Line Of Sight (LOS) one. This has been observed by ray tracing simulation which agreed also to the work presented in~\cite{Schafer1991} and~\cite{Wang2009}. Nevertheless, in most scenarios the positions of transmitting and receiving antennas are quasi-static with regard to each other. This allows using highly directive antennas, which significantly reduce the impact of multipath transmissions. Additionally, the antenna is circularly polarized so that the received power levels of the remaining multipath components are further reduced by mitigating the odd bounces of the multipath components. This high directive circularly polarized antenna will affect positively the channel impulse response by enhancing the power of LOS while mitigating the multipath components.\

A high directive circularly polarized antenna can be achieved using 4 * 4 patch antenna elements which are sequentially fed by equal amplitude and 90$^{\circ}$ phase shift between the adjacent elements. The receiving and transmitting antennas are designed identically to preserve the generality of the antenna subsystem. This design results in a high circular polarized directivity of ~18 dBi while the side lobe directivity is only ~4 dBi. Also, the odd bounces of multipath components will be received as cross-polarized resulting in a further reduction of the odd bounces multipath components.\

\section{Conclusions}
This paper presented a high level overview of the concept and architecture of a novel wireless technology for applications that require fast and reliable communication over short distances. The novelty of the proposed technology lies in combining the available state of the art methods that optimally exploit the available radio channel. Examples of such methods include using dedicated logical channels based on SP, simultaneous ranging and communication and the use of robust single carrier based air interface techniques.A demonstration system is also planned to be developed, which shows the possibilities under realistic operating conditions.

\bibliographystyle{IEEEtran}
\bibliography{SBDist_paper}

\section*{Acknowledgements}
Part of this work has been performed in the framework of the BMBF project SBDist. The authors would like to acknowledge the contributions of their colleagues, although the views expressed are those of the authors and do not necessarily represent the project.
\end{document}